\documentclass[twoside,11pt]{article}
\usepackage{pgm}

\usepackage[utf8]{inputenc}
\inputencoding{utf8}
\usepackage{amsfonts,amsmath,amssymb}
\usepackage{hyperref}
\usepackage{booktabs}
\usepackage{algorithm}
\usepackage{algorithmic}
\usepackage[normalem]{ulem}

\newcommand{\mat}[1]{\boldsymbol{\mathrm{#1}}}
\newcommand{\vecb}[1]{\boldsymbol{#1}}

\newcommand{\indep}{\perp \!\!\! \perp}
\newcommand{\given}{\mid}

\ShortHeadings{Simulating covariance and concentration graph matrices}{Córdoba et al.}

\begin{document}

\title{A partial
orthogonalization method for simulating covariance and concentration graph
matrices}

\author{\Name{Irene Córdoba$^{1}$}
\Email{irene.cordoba@upm.es} \\
        \Name{Gherardo Varando$^{{1},2}$}
        \Email{gherardo.varando@{upm.es}} \\
        \Name{Concha Bielza$^{1}$} \Email{mcbielza@fi.upm.es}\\
        \Name{Pedro Larrañaga$^{1}$} \Email{pedro.larranaga@fi.upm.es}\\
        \addr $^{1}$Department of Artificial Intelligence, Universidad Politécnica de Madrid (Spain) \\ 
    \addr $^{2}$Department of Mathematical Sciences, University of Copenhagen (Denmark)\thanks{Starting August 2018}}

\editor{}
\maketitle

\begin{abstract}%
Structure learning methods for covariance and concentration graphs are often
	validated on synthetic models, usually obtained by randomly generating: (i)
	an undirected graph, and (ii) a compatible symmetric positive definite (SPD)
	matrix. In order to ensure positive definiteness in (ii), a dominant
	diagonal is usually imposed. However, the link strengths in the resulting
	graphical model, determined by off-diagonal entries in the SPD matrix, are
	in many scenarios extremely weak. Recovering the structure of the undirected
	graph thus becomes a challenge, and algorithm validation is notably
	affected. In this paper, we propose an alternative method which overcomes
	such problem yet yielding a compatible SPD matrix. We generate a partially
	row-wise-orthogonal matrix factor, where pairwise orthogonal rows correspond
	to missing edges in the undirected graph. In numerical experiments ranging
	from moderately dense to sparse scenarios, we obtain that, as the dimension
	increases, the link strength we simulate is stable with respect to the
	structure sparsity. Importantly, we show in a real validation setting how
	structure recovery is greatly improved for all learning algorithms when
	using our proposed method, thereby producing a more realistic comparison
	framework.
\end{abstract}

\section{Introduction}\label{sec:intro}
Structure learning algorithms in graphical models are validated using
either benchmark or randomly generated synthetic models from which data is
sampled. This allows to evaluate
their performance by comparing the recovered graph,
obtained by running the algorithm over the generated data, with
the known true structure.  The synthetic graphical models are
typically constructed in a two-step manner: a graph structure is selected at
random or chosen so that it is representative of the problem at hand; and,
similarly, its parameters are fixed or randomly sampled.  

Covariance \citep{cox1993,kauermann1996} and concentration graphs
\citep{dempster1972,lauritzen1996} are graphical models where the variables are
assumed to follow a multivariate Gaussian distribution, and the structure is
directly read off in the covariance or concentration matrix, respectively.
Looking at the literature on these models, one finds that typical benchmark
structures are Toeplitz, banded, diagonally spiked and block diagonal covariance or
concentration matrices \citep{yuan2007,xue2012,ledoit2012}, with parameters
fixed to ensure positive definiteness. 

The issue of positive definiteness is specially relevant when the structure is
randomly generated. One approach to overcome this could be to sample from a
matrix distribution with support over the symmetric positive definite
matrices compatible with the undirected graph structure. The hyper Wishart
distributions \citep{dawid1993,letac2007} are the most well-developed in this
sense, since they form a conjugate family for Bayesian analysis. 
However, while sampling algorithms are available for general
concentration graphs \citep{carvalho2007,lenkoski2013}, in covariance graphs they
have been developed only in the decomposable case \citep{khare2011}. 

In general, hyper Wishart distributions are rarely used in validation scenarios
\citep{williams2018}, and instead in the literature the most common approach to
ensure positive definiteness is to enforce diagonal dominance in the covariance
or concentration matrix \citep{lin2009,arvaniti2014,stojkovic2017}. However,
when the undirected graph is moderately dense, the off-diagonal elements in the
generated matrices, often interpreted as link strengths, are extremely small
with respect to the diagonal entries and structure recovery becomes a
challenge, thereby compromising the structure learning algorithm validation
\citep{schafer2005,schafer2005b,kramer2009,cai2011}.
In this paper, we propose an alternative method to overcome this problem based
on partial orthogonalizations. In particular, we build a matrix factor where
pairwise orthogonal rows correspond to missing edges in the undirected graph.
Our method does not suffer from the problem of weak link strengths, as we
numerically check in a wide range of sparsity scenarios. We also use our
simulation method in a real validation setting and show how the performance is
greatly improved for every learning algorithm, thereby potentially changing the
conclusions drawn if only using diagonally dominant matrices for comparison.

The rest of the paper is organized as follows. Preliminaries are introduced in
Section \ref{sec:prel}, where we briefly overview concentration and covariance
graphs, and the main characteristics of diagonally controlled matrices. Next, in
Section \ref{sec:methods}, we present our partial orthogonalization method,
analyzing its main properties and our particular implementation.
Section \ref{sec:exp} contains a description of the experiment set-up we have
considered, and the interpretation of the results obtained. Finally, in Section
\ref{sec:conc} we conclude the paper and outline our plans for future
research.

\section{Preliminaries}\label{sec:prel}
In the remainder of the paper, we will use the following notation. We let $X_1,
\ldots, X_p$ denote $p$ random variables and $\vecb{X}$ the random vector they
form.  For each subset $I \subseteq \{1, \ldots, p\}$, $\vecb{X}_I$ will be the
subvector of $\vecb{X}$ indexed by $I$, that is, $(X_i)_{i \in I}$. We
follow~\cite{dawid1980} and abbreviate conditional independence in the joint
distribution of $\vecb{X}$ as $\vecb{X}_I \indep \vecb{X}_J \given \vecb{X}_K$,
meaning that $\vecb{X}_I$ is conditionally independent of $\vecb{X}_J$ given
$\vecb{X}_K$, with $I, J, K$ pairwise disjoint subsets of indices. Entries in a
matrix are denoted with the respective lower case letter, for example, $m_{ij}$
denotes the $(i, j)$ entry in matrix $\mat{M}$. 

\subsection{Gaussian graphical models}

Covariance and concentration graphs are graphical models where it is assumed
that the statistical independences in the distribution of a multivariate
Gaussian random vector $\vecb{X} = (X_1, \ldots, X_p)$ can be represented by an
undirected graph $G = (V, E)$. Typically, $\vecb{X}$ is assumed to have zero
mean for lighter notation, and $V = \{1, \ldots, p\}$ so that it indexes the
random vector, that is, $\vecb{X}_V = \vecb{X}$. We will represent the edge set
$E$ as a subset of $V \times V$, therefore $(i, j) \in E$ if and only if $(j, i) \in E$.

In covariance graphs, the independences represented are marginal, meaning that
whenever there is a missing edge $(i,j)$ in $G$, the random
variables $X_i$ and $X_j$ are marginally independent. More formally, this is
called the pairwise Markov property of covariance graphs
\citep{cox1993,kauermann1996},
\[
	X_i \indep X_j \quad \text{for } i,j \in V \text{ s.t. } i \not\sim_{G} j, 
\]
where $i \sim_{G} j$ is the adjacency relationship on the graph $G$, that is, $i
\sim_{G} j$ if and only if $(i,j) \in E$. Note further that $X_i \indep X_j$ if
and only if $\sigma_{ij} = 0$.

By contrast, in concentration graphs, a missing edge implies a conditional
independence; specifically, in this case the pairwise Markov property
\citep{lauritzen1996} becomes
\[
	X_i \indep X_j \given \vecb{X}_{V \setminus  \{i, j\}}
        \quad \text{for } i,j \in V \text{ s.t. } i \not\sim_{G} j. 
\]
In turn, this can be read off in the concentration matrix $\mat{\Omega} =
\mat{\Sigma}^{-1}$, that is, $X_i \indep X_j \given \vecb{X}_{V \setminus \{i,
j\}} \iff \omega_{ij} = 0$.

One can always construct multivariate Gaussian distributions belonging to a covariance
or concentration graph, for an arbitrary structure $G$. Furthermore, these
models are Markov equivalent, in the sense that they represent the same set of
distributions, whenever the respective structures share the same disconnected
complete subgraphs \citep{jensen1988,drton2008}. This implies that the
statistical independences in most multivariate Gaussian distributions can only
be represented by either a covariance or a concentration graph. 

\subsection{Symmetric positive definite matrices and undirected graphs} 

The statistical independences implied by both {covariance} and {concentration}
graph models are explicitly represented in a symmetric positive definite matrix.
It is of our interest the problem on how to simulate such kind of matrices,
subject to the constraint of being compatible with a given undirected graph. We
will abstract ourselves from whether such graph has been randomly generated or
pre-specified.

Denote as $\mathbb{S}$ the space of symmetric {$p \times p$} matrices {and as
$\mathbb{S}^{>0}$ its subspace of symmetric positive definite matrices}. For a
fixed undirected graph $G$ let $\mathcal{M}_G$ be the set of matrices $\mat{M}$
with zeros in the entries represented by the missing edges in $G$, that is, 
\[
	\mathcal{M}_{G} = \{\mat{M} \in {\mathbb{R}^{p\times p}}
	\text{ s.t. } m_{ij} = m_{ji} = 0 \text{ if } i
	\not\sim_{G} j   \}.
\]
Let $\mathbb{S}(G) = \mathbb{S} \cap \mathcal{M}_{G}$ and  $\mathbb{S}^{>0}(G) =
\mathbb{S}^{>0} \cap \mathcal{M}_{G} $ be the sets of symmetric and symmetric
positive definite matrices with undirected graphical constraints.

Note that the covariance matrix $\mat{\Sigma}$ of a Gaussian random vector
$\vecb{X}$ whose distribution belongs to a covariance graph with structure $G$ satisfies that
$\mat{\Sigma} \in {\mathbb{S}^{>0}(G)}$. Analogously, if the distribution
belongs to a concentration graph with structure $G$, then $\mat{\Omega} =
\mat{\Sigma}^{-1} \in {\mathbb{S}^{>0}(G)}$. In either case it is clear that the
goal is to simulate elements belonging to $\mathbb{S}^{>0}$.

\subsection{Diagonally Controlled Matrices}
\label{sec:diagcontrolled}
When a matrix $\mat{M} \in \mathbb{S}$ satisfies that $m_{ii} > \sum_{j\neq i}
|m_{ij} |$ for each $i \in \{1, \ldots, p\}$, then $\mat{M}$ belongs to
$\mathbb{S}^{>0}$. Thus a simple method to generate a matrix in
{$\mathbb{S}^{>0}(G)$} consists in generating a random matrix {in
$\mathbb{S}(G)$} and then choosing diagonal elements so the final matrix is
diagonally dominant, as in Algorithm \ref{alg:domdiag}. The usual approach for
generating the initial matrix in line \ref{alg:domdiag:mat} is to use
independent and identically distributed (i.i.d.) 
nonzero entries. The diagonal dominance method has been
extensively {used} in the literature mainly for its simplicity and the ability
to control the singularity of the generated matrices, as we will now explain. 

It is possible to control the minimum eigenvalue of a matrix by varying its
diagonal elements \citep{honorio12}. In particular, let $G$ be an undirected
graph, $\mat{M}$ a matrix in $\mathbb{S}(G)$, and $\epsilon > 0$ the desired
lower-bound on the eigenvalues.
If $\lambda_{min}$ is the minimum eigenvalue of $\mat{M}$, then $\mat{M} +
(\lambda_{min}^{-} + \epsilon)\mat{I}_{p}$ belongs to $\mathbb{S}^{>0}(G)$ and
has eigenvalues greater or equal to $\epsilon$, where $\lambda_{min}^{-}$
denotes the negative part of $\lambda_{min}$.

Similarly, one can control the condition number, with respect to the Frobenious
norm, of the generated matrix as follows~\citep{cai2011}. If $\kappa_0 > 1$ is
the desired condition number and we already have a matrix $\mat{M} \in {\mathbb{S}(G)}$ with maximum
eigenvalue $\lambda_{max}>0$, then
\[
	\mat{M} + \frac{\lambda_{max} - \kappa_0\lambda_{min}}{\kappa_0 - 1} \mat{I}_{p}
\]
belongs to $\mathbb{S}^{>0}(G)$ and has condition number equal to $\kappa_0$.
Covariance and concentration matrices with an upper bound on the condition
number are attractive in certain estimation scenarios \citep{joongho2013}.

\begin{algorithm} 
	\caption{Simulation of a matrix in $\mathbb{S}^{>0}(G)$ using diagonal 
        dominance}\label{alg:domdiag}
	\begin{algorithmic}[1]

	\REQUIRE Undirected graph $G$
    \ENSURE Matrix belonging to $\mathbb{S}^{>0}(G)$
	
		\STATE $\mat{M} \gets$ random $p \times p$ matrix in $\mathbb{S}(G)$
		\label{alg:domdiag:mat}
		\FOR{$i = 1, \ldots, p$}
			\STATE $m_{ii} \gets \sum_{i \neq j} | {m_{ij}} | +$ random positive
			perturbation
		\ENDFOR
		
		\RETURN $\mat{M}$
	\end{algorithmic} 
\end{algorithm}

\section{Simulating matrices in {$\mathbb{S}^{>0}(G)$} by partial orthogonalization}\label{sec:methods}
Let $G_p$ be an Erdos-Renyi~\citep{erdos1959} random graph over $p$ nodes with
edge probability $d\in (0,1)$ and given $G_p$ let $\mat{M} \in \mathbb{S}(G_p)$
be a symmetric random matrix with i.i.d. non-zero
off-diagonal entries $m_{ij}$ following a distribution with  $0< \mu < +\infty$
the expected absolute value.  If we denote with $\mat{M}' \in
\mathbb{S}^{>0}(G_p)$ the output of the diagonal dominance method
(Section~\ref{sec:diagcontrolled}), we have that for all $1\leq i \leq p$ and $j
\neq i$: 

\begin{equation} \label{eq:ratio} r_{ij} = \frac{|m'_{ij}|}{m'_{ii}}< 
\frac{|m_{ij}|}{\sum_{t\neq i} |m_{it}|} <  \frac{|m_{ij}|}{\sum_{t\neq
    i,j } |m_{it}|}= 
    \frac{|m_{ij}|}{(p-2) \sum_{t\neq i,j} |m_{it}| /
    (p-2)}.
\end{equation}
By the strong law of large number and since $|m_{ij}|$ and $\sum_{t\neq
i,j}|m_{it}|$ are independent, we have that $r_{i,j}  \xrightarrow[]{a.s.} 0$.

In order to overcome such issue, we propose an alternative method which doesn't
rely on diagonal dominance, which we will now describe. If we consider an
arbitrary $p \times p$ full rank matrix $\mat{Q}$, the product $\mat{Q}\mat{Q}^{t}$ is positive
definite and symmetric, and therefore lies in $\mathbb{S}^{>0}$. Moreover, $\mat{Q}\mat{Q}^t$ belongs to
$\mathbb{S}^{>0}(G)$ if and only if 
\[
	\vecb{q}_i \bot \vecb{q}_j \quad \text{for } i \not \sim_{G} j,
\]
where $\bot$ denotes orthogonality with respect to the standard scalar product
on $\mathbb{R}^p$, and $\vecb{q}_i$ is the $i$-th row of $\mat{Q}$.  

Thus, given an undirected graph $G$, we can impose Markov properties for the matrix
$\mat{Q}\mat{Q}^t$ simply by orthogonalizing the respective rows of $\mat{Q}$.
The pseudocode for the described procedure can be found in Algorithm
\ref{alg:partort}. 

\begin{algorithm} 
	\caption{Simulation of a matrix in $\mathbb{S}^{>0}(G)$ using partial
        orthogonalization}\label{alg:partort}
	\begin{algorithmic}[1]

	\REQUIRE Undirected graph $G$
    \ENSURE Matrix belonging to $\mathbb{S}^{>0}(G)$
	
		\STATE $\mat{Q} \gets$ random $p \times p$ matrix
		\FOR{$i = 1, \ldots, p$}
			\STATE orthogonalize $\vecb{q}_i$ with respect to the span of
						$\{ \vecb{q}_j \text{ s.t. } i
						\not\sim_G j \text{ and } j < i  \} $
		\ENDFOR
		
		\RETURN $\mat{Q}\mat{Q}^t$\label{alg:partort:ret}
	\end{algorithmic} 
\end{algorithm}

\begin{figure}[h]
\centering
\includegraphics[scale = 0.4]{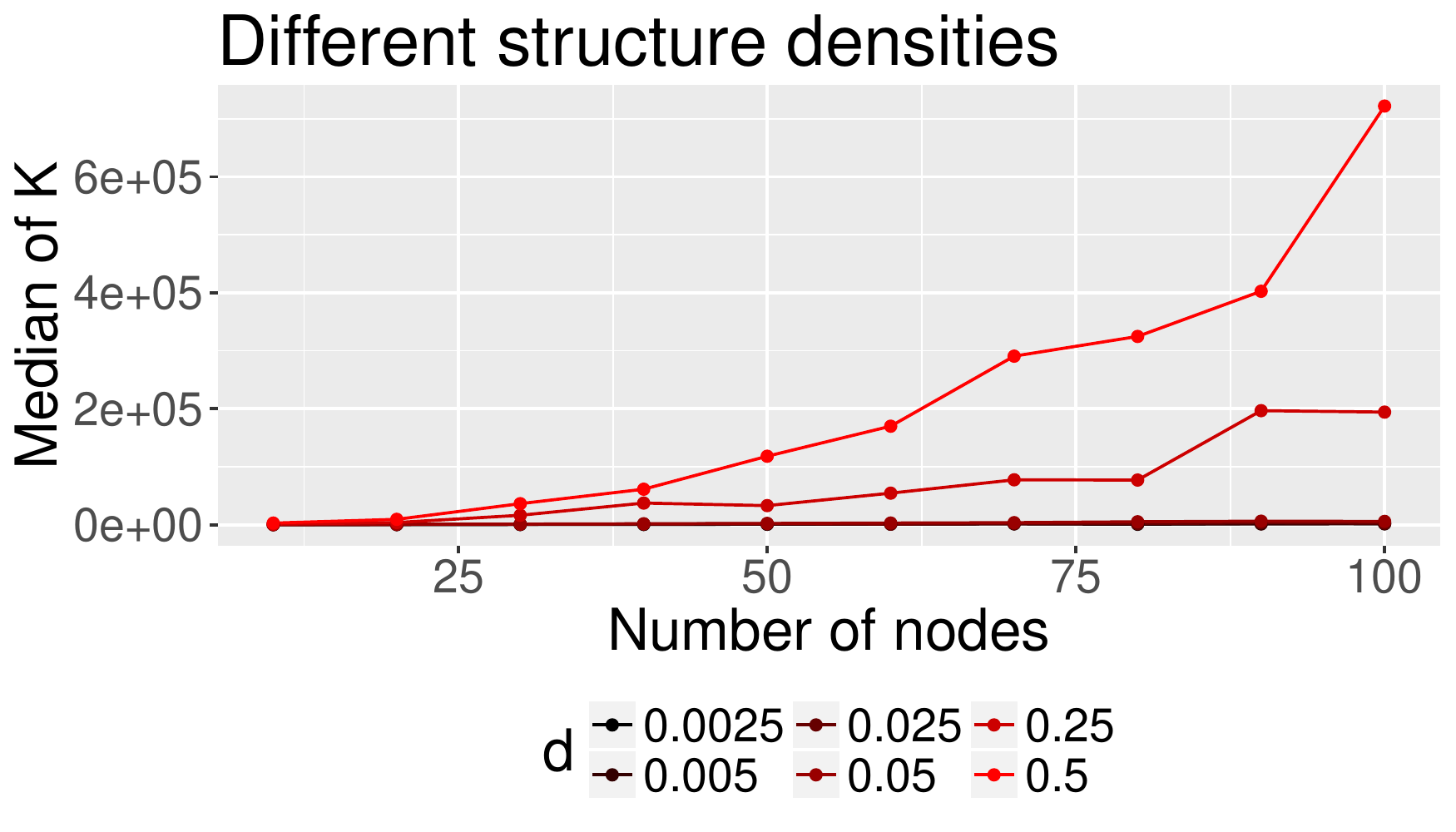} 
\includegraphics[scale = 0.4]{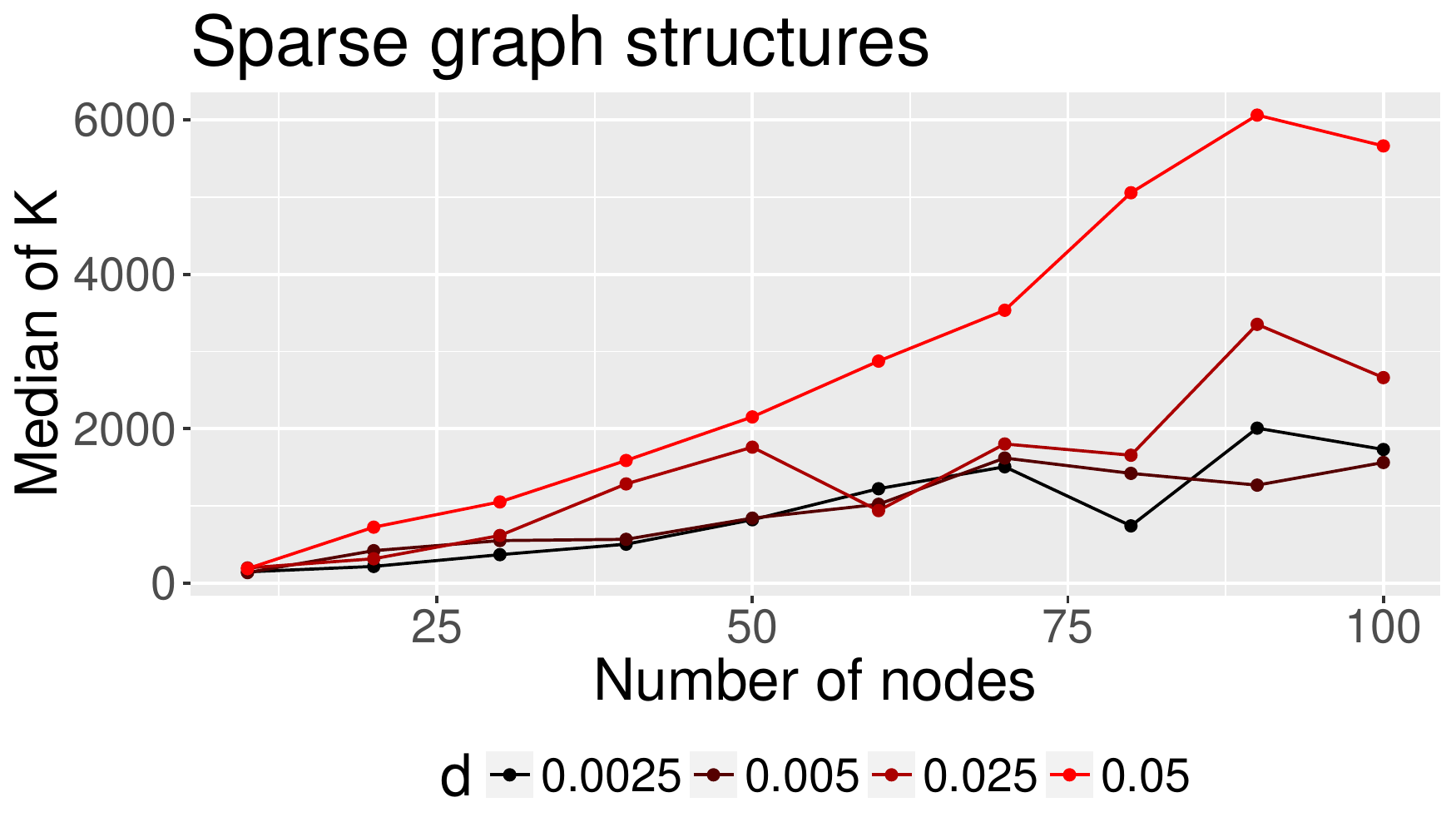} 
    \caption{The median of the condition number as a function of the number of
		variables $p$ for different structure densities $d$. We can see that the
		lower bound of $p^2$ is valid for dense structures ($d = 0.5$ and $d =
		0.25$ in the left), whereas sparse graphs yield lower condition numbers
		(right figure). K: condition number
	of the matrix.}
\label{fig:kappa} 
\end{figure}

After Algorithm \ref{alg:partort} has finished, it outputs a
matrix that correctly reflects the graphical structure given by the input graph
$G$. If the entries in matrix $\mat{Q}$ are initially simulated as i.i.d.
centered subgaussian, then its condition number $\kappa(\mat{Q}) \geq p$ with high
probability \citep{rudelson2009}.
Therefore, in such case the condition number of the matrices $\mat{Q}\mat{Q}^t$
returned by Algorithm \ref{alg:partort} will satisfy $\kappa(\mat{Q}\mat{Q}^t) \geq
p^2$ as the graph structure becomes denser, as shown in Figure \ref{fig:kappa}.
Although the magnitude of the condition numbers shown are relatively high, this
has not been an issue in our numerical experiments (see Section \ref{sec:exp}).

In particular we can use a modified Gram-Schmidt orthogonalization procedure
that iteratively orthogonalizes every row $\vecb{q}_i$ with {respect to} the
set of rows $i^\bot = \{\vecb{q}_j \text{ s.t. }i \not \sim_{G} j \text{ and } j < i \}$.
This particularization of the proposed method is reflected in Algorithm
\ref{alg:partortgs}, where $proj_{\vecb{v}}(\vecb{u})$ denotes the orthogonal
projection of a vector $\vecb{u}$ on another vector $\vecb{v}$. The loop in line \ref{alg:partort:vsel}
constructs a set of orthogonal vectors $\vecb{\tilde{q}}_j$ which span the same subspace
than the original rows $\vecb{q}_j$ belonging to $i^\bot$. This orthogonal base
is later used in the loop at line \ref{alg:partort:ort} for ensuring that
$\vecb{q}_i$ is jointly orthogonal to all the vectors in $i^\bot$. Note however
that these auxiliary orthogonal set of vectors is discarded in the next
iteration, and the original $\vecb{q}_j$ are kept in the factor matrix $\mat{Q}$
that will be used in the last computation of line \ref{alg:partort:return}.

\begin{algorithm} 
	\caption{Simulation of a matrix in $\mathbb{S}^{>0}(G)$ by modified
	Gram-Schmidt orthogonalization}\label{alg:partortgs}
	\begin{algorithmic}[1]

	\REQUIRE Undirected graph $G$
            \ENSURE Matrix belonging to
            {$\mathbb{S}^{>0}(G)$}
		\STATE $\mat{Q} \gets$ random $p \times p$ matrix
		\FOR{$i = 1, \ldots, p$}
			\FOR{$j = 1, \ldots, i - 1$}
			\label{alg:partort:vsel}
					\IF{$i \not \sim_G j$} 
						\STATE $\vecb{\tilde{q}}_j \gets \vecb{q}_j$
						\FOR{$k = 1, \ldots, j - 1$}
							\STATE $\vecb{\tilde{q}}_j \gets \vecb{\tilde{q}}_j -
							proj_{\vecb{\tilde{q}}_k}(\vecb{\tilde{q}}_j)$
						\ENDFOR
					\ENDIF
			\ENDFOR
			\FOR{$j = 1, \ldots, i - 1$}
			\label{alg:partort:ort}
            	\STATE $\vecb{q}_i \gets \vecb{q}_i -
				proj_{\vecb{\tilde{q}_j}}(\vecb{q}_i) $
			\ENDFOR
		\ENDFOR
		
		\RETURN $\mat{Q}\mat{Q}^t$\label{alg:partort:return}
	\end{algorithmic} 
\end{algorithm}
	
The computational complexity of Algorithm~\ref{alg:partortgs} is mainly given by
the loop in line~\ref{alg:partort:vsel}, where an orthogonal base is found for
the subspace spanned by nonadjacent rows to $i$. In the worst case scenario,
such row set $i^\perp$ has cardinality $i - 1$, becoming the complexity of the
loop $O(i^2p)$, giving an overall worst case complexity $O(p^4)$.

\section{Numerical experiments}\label{sec:exp}
In this section we perform a simulation study to compare our proposed method to
generate matrices in {$\mathbb{S}^{>0}(G)$} against the diagonal dominance one.
We have used random Erd\"os-R\'enyi{~\citep{erdos1959}} undirected graphs $G =
(V, E)$. The size of the vertex set $V$, $p$, will take each of the values in
the first row of Table \ref{tab:exp}. The probability of the inclusion of an
edge in $E$, $d$, will take the values displayed in the second row of Table
\ref{tab:exp}. This probability can be thought of as an indicator of the graph's
density, ranging from sparse structures ($d = 0.0025$) to dense ones ($d =
0.5$). In particular, for every $(p,d) \in P\times D$ (Table \ref{tab:exp}) we
generate $10$ Erd\"os-R\'enyi graphs, $G^{p,d}_1, \ldots, G^{p,d}_{10}$, and we
sample $10$ matrices in $\mathbb{S}^{>0}(G^{p,d}_n)$ ($n \in \{1, \ldots, 10\}$)
using our proposed method (Algorithm \ref{alg:partort}) and diagonal dominance
(Algorithm \ref{alg:domdiag}). In total we thus sample $100$ matrices for every pair of
parameters $(p,d) \in P\times D$.  Both methods need to generate a matrix with
random entries as a first step. In order to generate the initial matrices in
both methods, we sample i.i.d. entries following a uniform distribution
$\mathrm{U}[0, 1]$. 

\begin{table}[h!]
	\centering
	\begin{tabular}{l l}		
		\toprule
		Parameter & Value set\\
		\midrule
		$p$ & $P = \{10, 20, \ldots, 100,125,150,200,250,300,400,500,750,1000\}$ \\
		$d$ & $D = \{0.0025, 0.005, 0.025, 0.05, 0.25, 0.5\}$ \\
		\bottomrule
	\end{tabular}
	\caption{Setting of the numerical experiments for simulating from
		$\mathbb{S}^{>0}(G)$, with $G$ an undirected graph. Values for
		the size of the vertex set ($p$) and the density of the structure ($d$).}
	\label{tab:exp}
\end{table}

\begin{figure}[h]
\centering
\includegraphics[scale = 0.4]{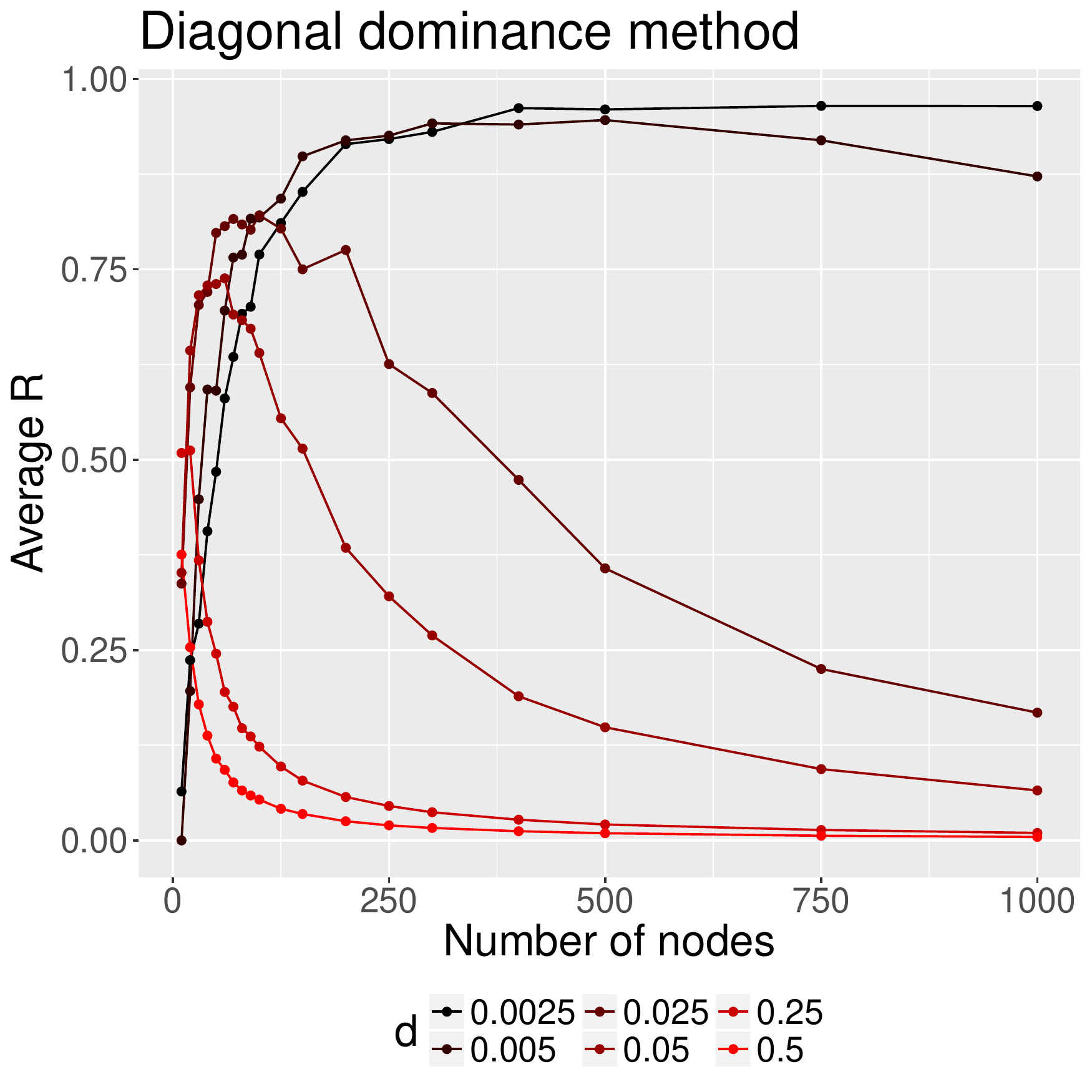} 
\includegraphics[scale = 0.4]{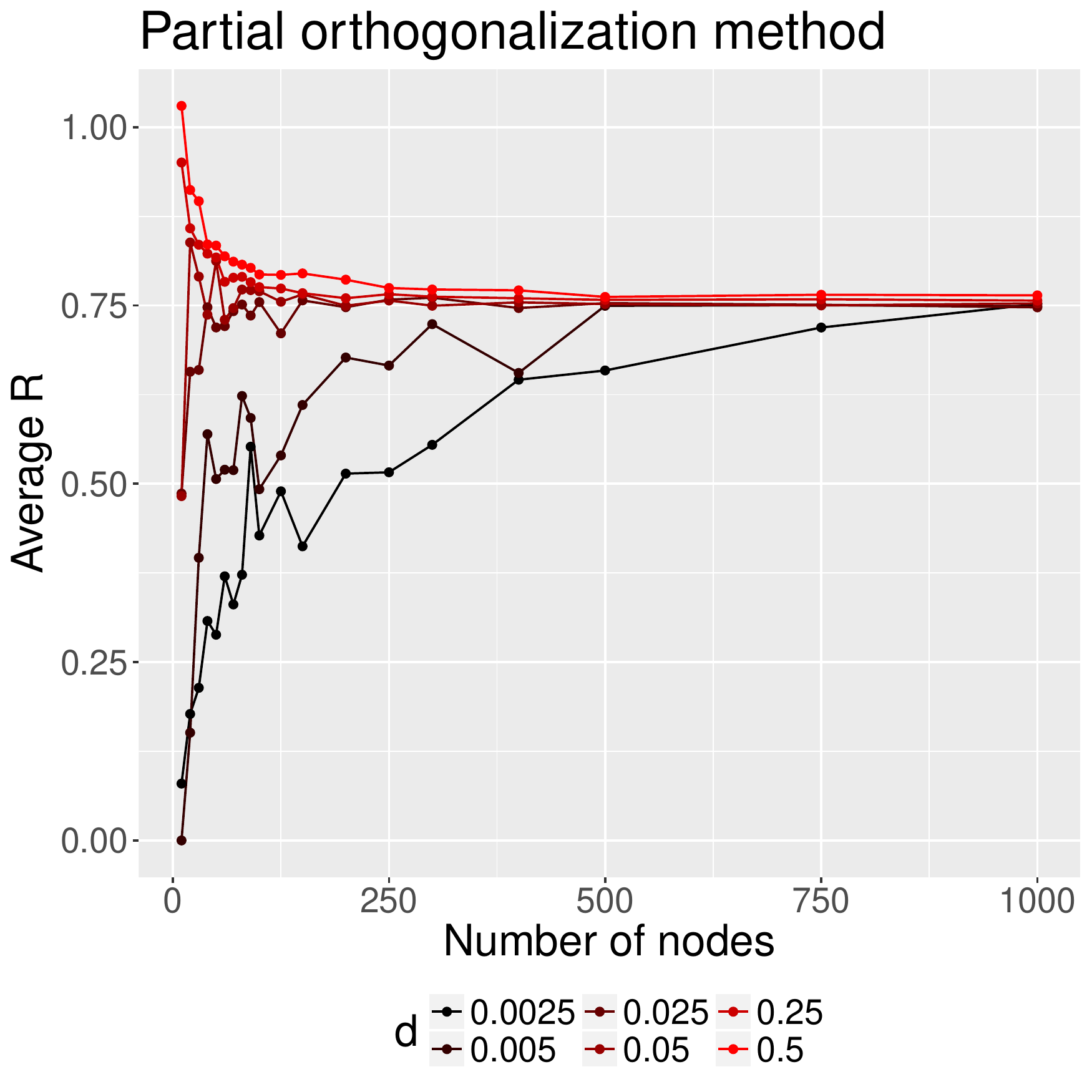} 
    \caption{The average of $R$ as a function of the number of
		variables $p$ for different structure densities $d$.}
\label{fig:avgmaxratio} 
\end{figure}

We compute for every $(p, d)$ in Table \ref{tab:exp} the
average $\overline{R}$ of the maximum ratio $R = \max_{j \neq i}
r_{ij}$, where $r_{ij}$ are defined as in Equation~\eqref{eq:ratio} and the
dependence on the matrix under consideration has been omitted for notational
simplicity.
A plot of the behaviour of $\overline{R}$ as a function of the number
of variables in the model, for different density values of the graphical
structure, is shown in Figure~\ref{fig:avgmaxratio}. 
We can observe that our proposed method generates matrices with an
asymptotically (in the number of variables $p$) constant value of
$\overline{R}$. On the contrary for undirected graphs whose density is higher
than $0.025$, which are usually found in applications \citep{kramer2009}, $\overline{R}$
goes to zero as $p$ increases for matrices simulated using the diagonal
dominance method. Only for arguably very
sparse matrices ($d \leq 0.005$), this method is able to avoid
such asymptotic behaviour with respect to $p$, but as we have shown in Section
\ref{sec:methods} for sufficiently high values of $p$ the same behaviour is to
be expected.
In particular, since the $\operatorname{U}[0,1]$ distribution is bounded, we can obtain
from Equation~\ref{eq:ratio} that almost surely $r_{ij} \leq 
2(p-1)^{-1} d^{-1}$ asymptotically and thus $R = \mathcal{O}(p^{-1})$, thereby
allowing for an approximate computation of the number of variables $p_0$ from where,
for a given structure density $d_0$, $R$ is arbitrarily close to zero.

\begin{figure}[h]
\centering
\includegraphics[scale = 0.4]{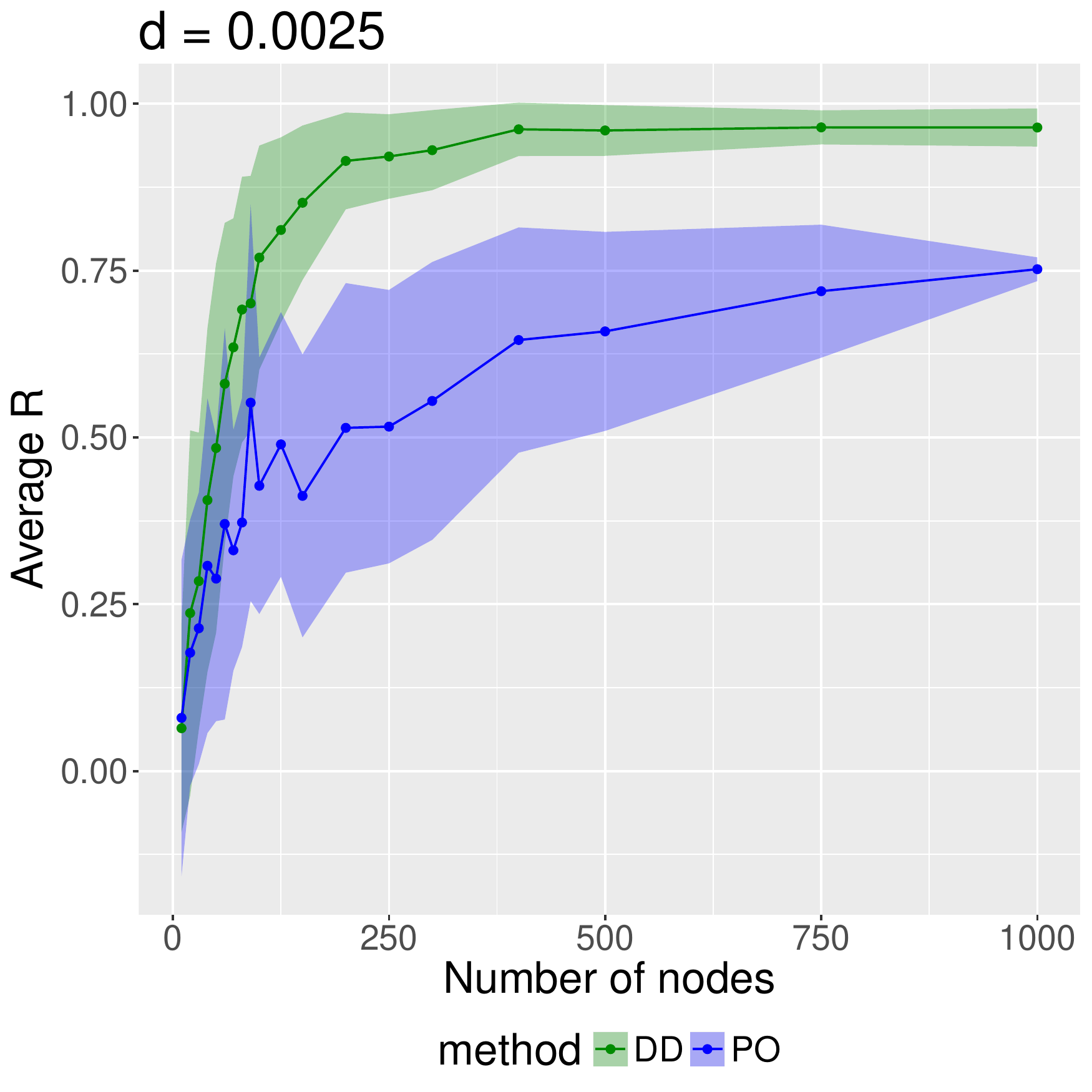}
\includegraphics[scale = 0.4]{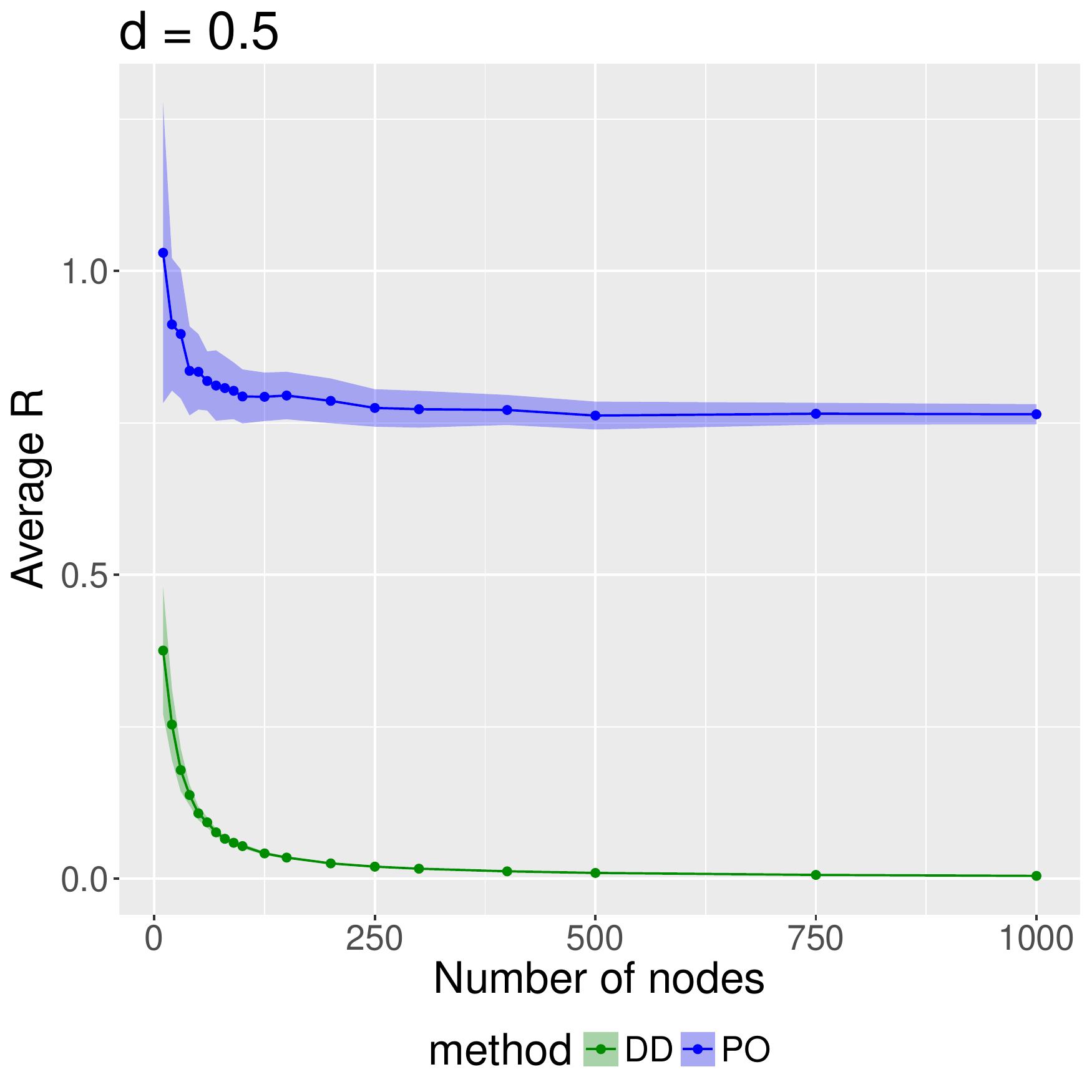}
    \caption{The average of $R$ as a function of number of
		variables $p$, for the two extreme value{s} of $d${:} very sparse matrices (left) and
    very dense matrices (right). Standard deviation of the {average} is shown.
DD: Diagonal dominance method; PO: Partial orthogonalization method.}
\label{fig:cmpavgmaxratio}
\end{figure}

The above mentioned conclusions are complementarily drawn from Figure
\ref{fig:cmpavgmaxratio}, where we have jointly plotted the performance of both
methods for the two extreme values we have considered for the structure density:
$0.0025$ (very sparse) and $0.5$ (very dense). We also show as a shade one
standard deviation on either side of the mean.  We can observe that in the sparse
scenario both methods perform reasonably good, with the diagonal dominance
method being more stable in terms of the standard deviation; however, we must
also point out that as $p$ increases, our method becomes more robust, being
equally stable for structures of a thousand vertices. By contrast, in the dense
case the diagonal dominance method performance is terribly affected early, being
almost zero for $p > 125$, approximately. Our proposed method, however, manages
to achieve a reasonable value for the average ratio, and the constant behaviour in
$p$ can be clearly observed. We also obtain that our method is more stable for
dense structures than in the sparse case. In some sense, this is not surprising,
because the more missing edges in the undirected graph of the model, the more
orthogonalizations in Algorithm \ref{alg:partort}, therefore the numerical
stability is more compromised, even though competitive results are equally
obtained.

\begin{figure}[h!]
\centering
\includegraphics[scale = 0.4]{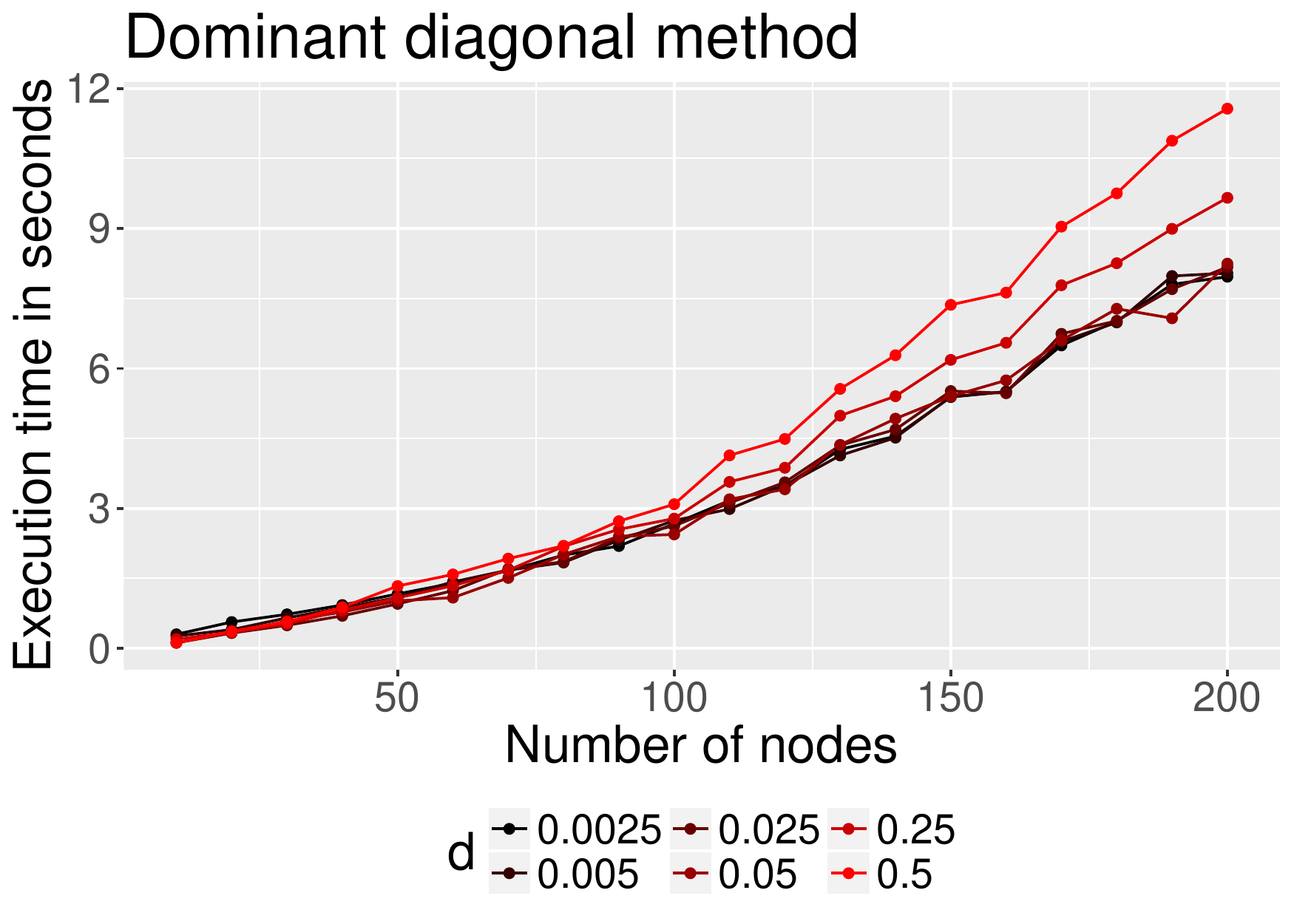}
\includegraphics[scale = 0.4]{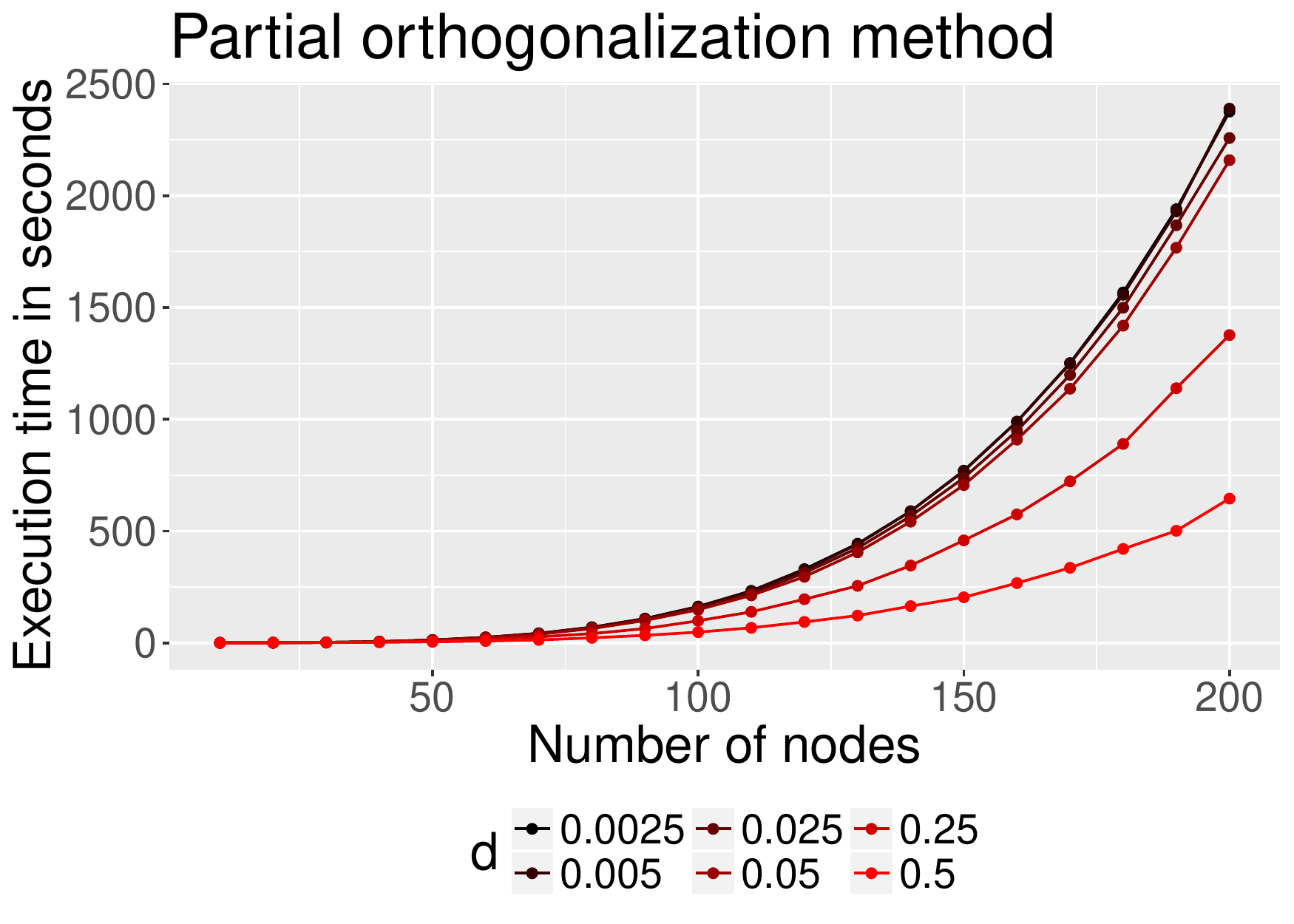}
\caption{Execution time to simulate $5000$ matrices.}
\label{fig:time}
\end{figure}

We have also measured the execution time of the two approaches. For this, we
have sampled $5000$ matrices for the different density values in Table
\ref{tab:exp}, and for a number of variables ranging from $10$ to $200$, in
different step sizes. This experiment has been executed on a machine equipped
with Intel Core i7-5820k, $3.30$ GHz$\times12$ and $64$ GB of RAM.  The results
are shown in Figure~\ref{fig:time}. The diagonal dominance
method is few orders of magnitude faster than our proposed method, which is
somewhat expected given its relative simplicity. We observe how the
computational cost of the partial orthogonalization method depends on the
structure density. For small values of $d$ the undirected graph contains a lot
of disconnected vertices and thus we repeat the loop in line
\ref{alg:partort:vsel} of Algorithm \ref{alg:partortgs} for many matrix rows,
being closer to the worst case scenario of $\mathcal{O}(p^4)$. In
practice, however, when validating structure learning algorithms only few
matrices need to be generated, which is completely affordable in all scenarios of 
Figure \ref{fig:time}. For example, the time needed to generate 10 matrices with
200 nodes in the worst case is approximately five seconds.

\begin{figure}[h!]
\centering
\includegraphics[scale = 0.4]{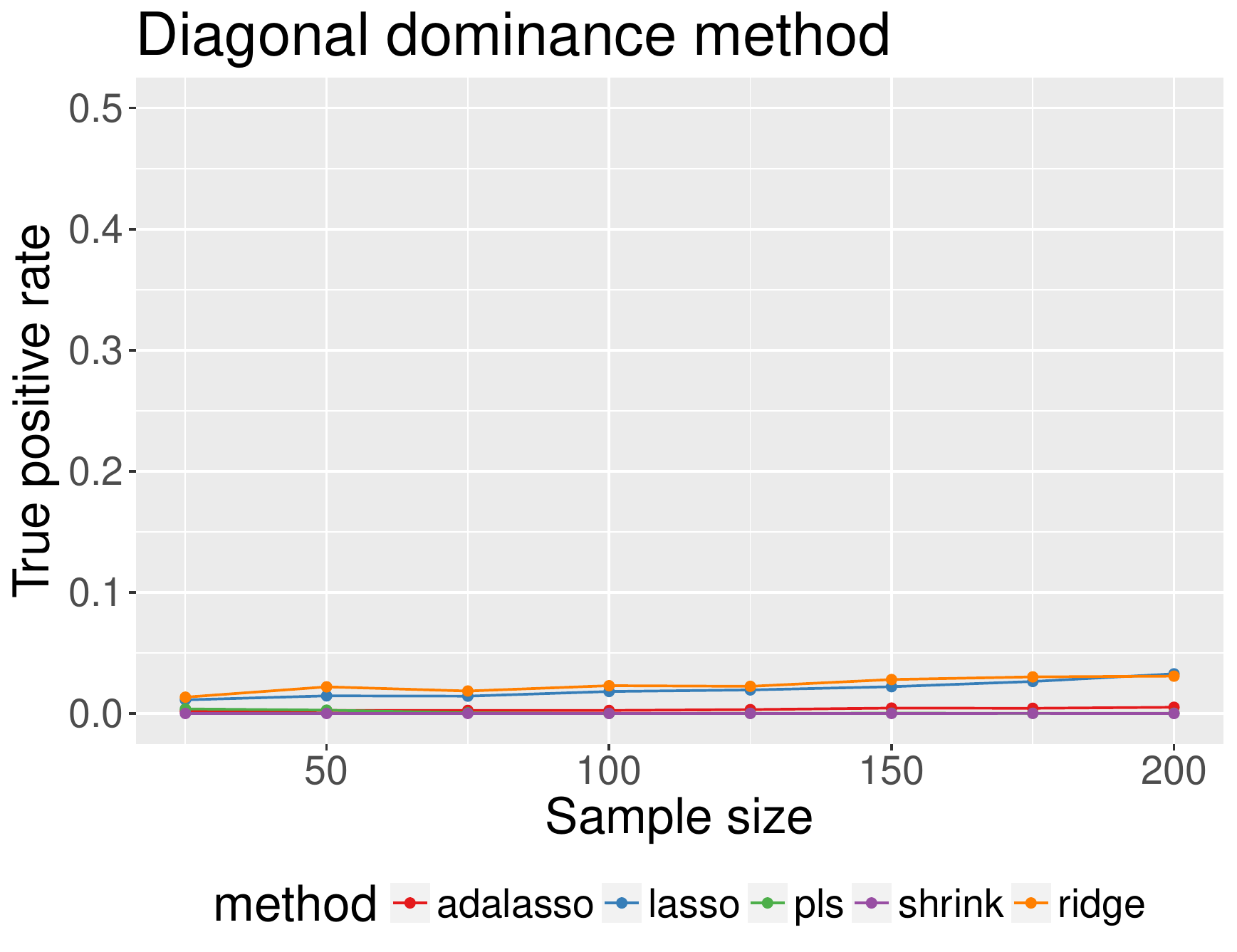}
\includegraphics[scale = 0.4]{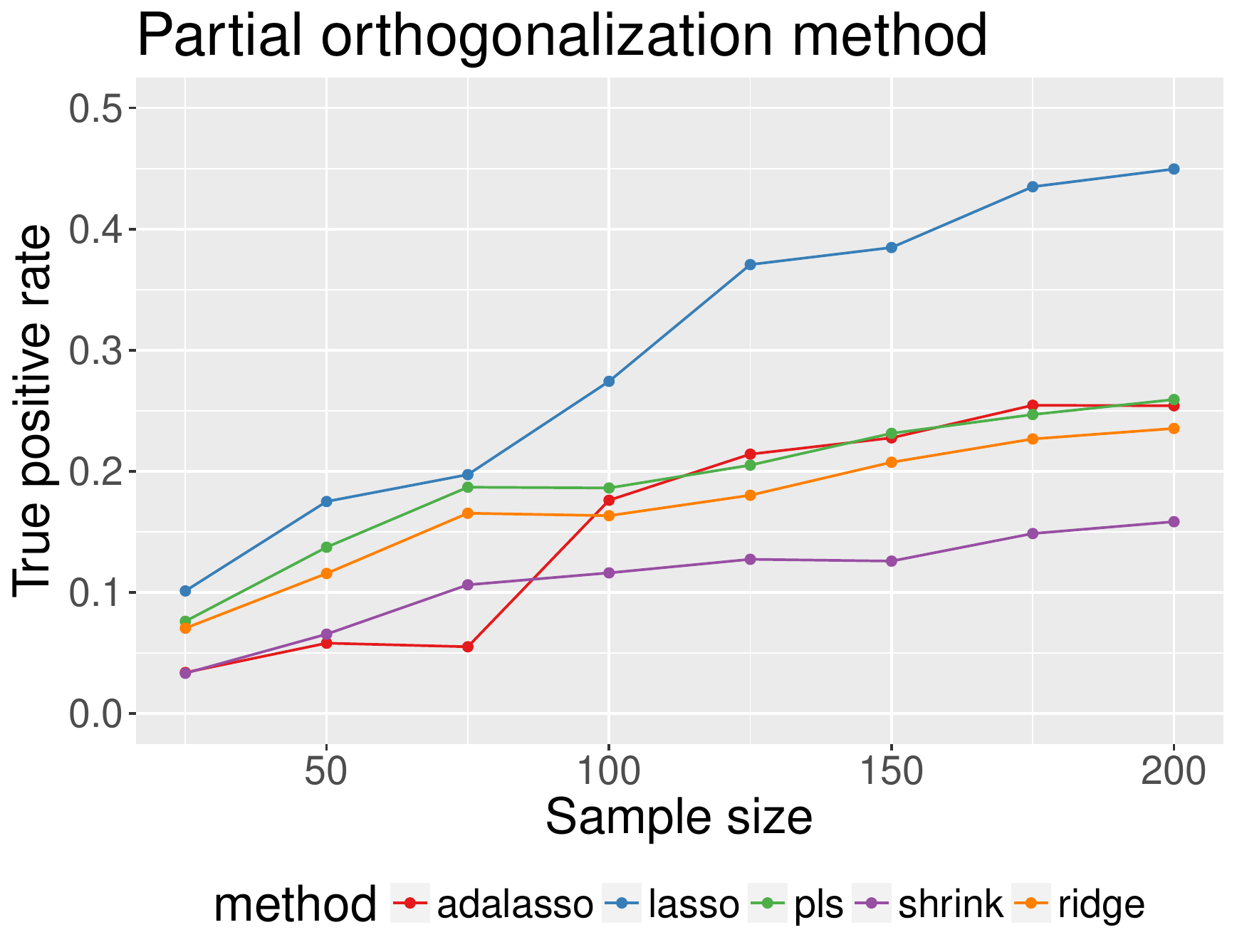}
\includegraphics[scale = 0.4]{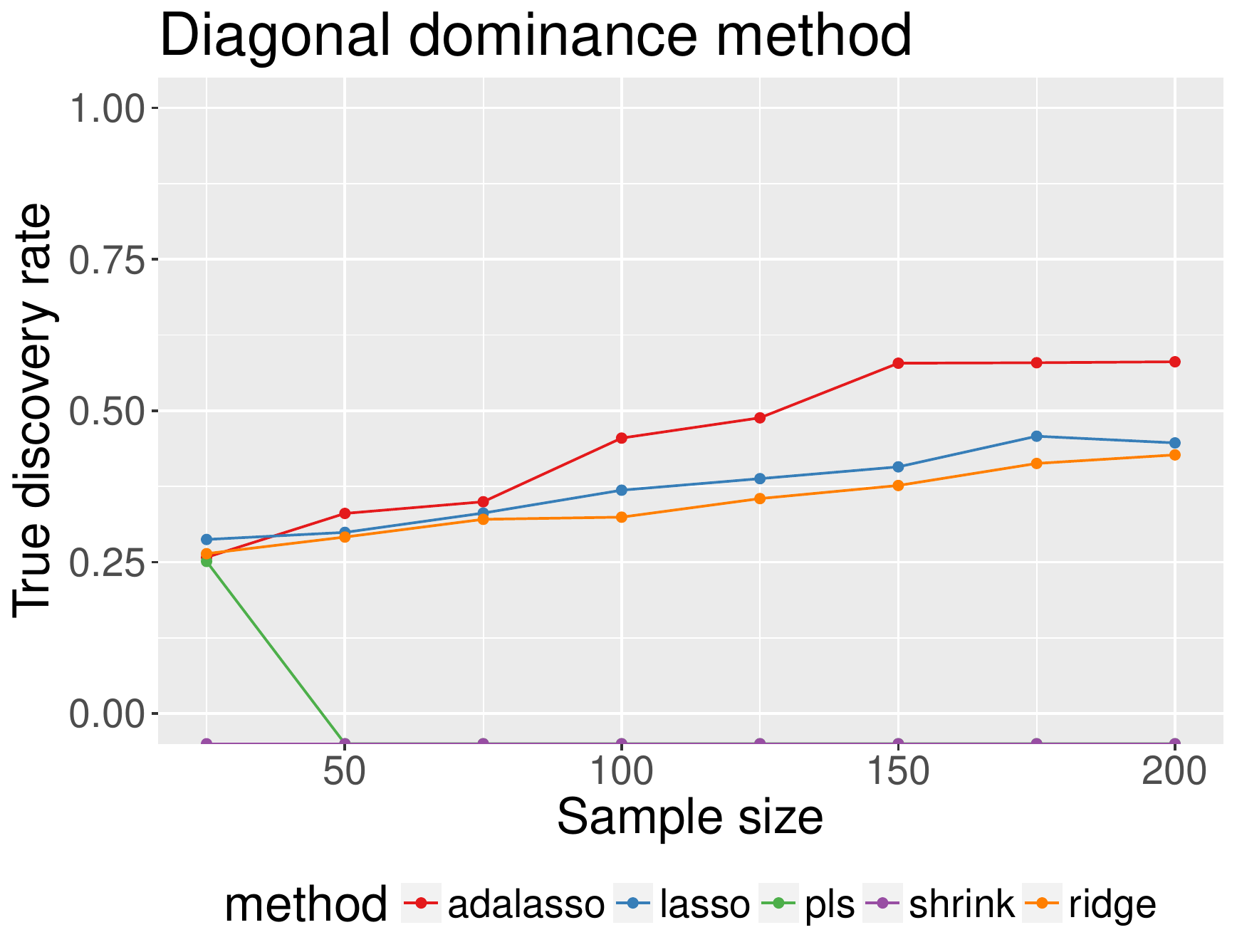}
\includegraphics[scale = 0.4]{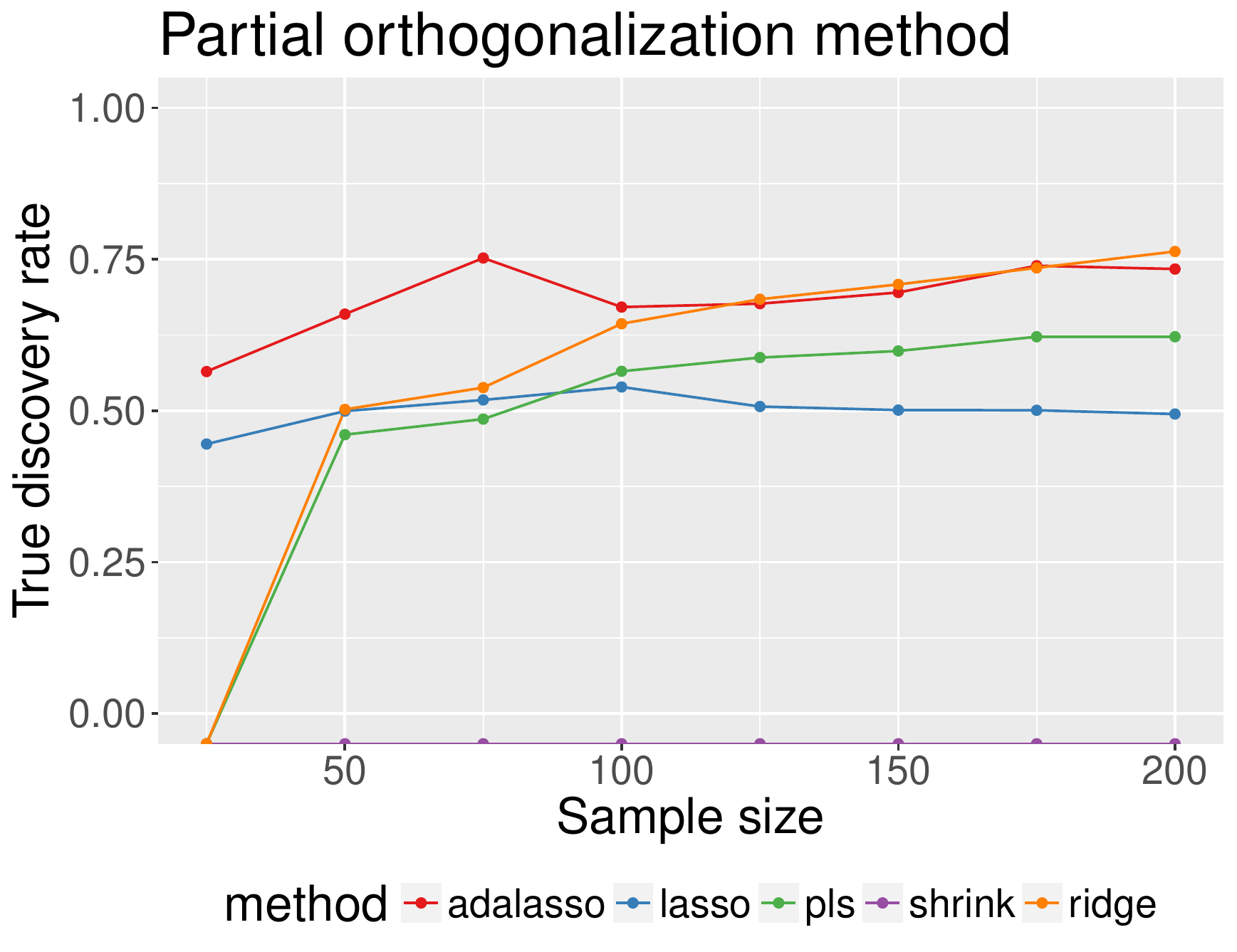}
\caption{True positive rate and true discovery rate of the structure learning
	algorithms for concentration graphs validated in \citep{kramer2009}. The
	number of variables (vertices in the undirected graph and dimension of the
	generated matrices) is fixed at $100$. \texttt{adalasso}: Adaptive $l_1$
regularization; \texttt{lasso}: $l_1$ regularization; \texttt{pls}: partial
least squares regression; \texttt{shrink}: shrinkage estimator of
\citet{schafer2005b}; \texttt{ridge}: $l_2$ regularization.}
\label{fig:kramer}
\end{figure}

The main motivation for the proposed method are the observations
that can be found in the literature on covariance and concentration graphs
regarding the difficulties of validating the performance of
structure learning algorithms \citep{schafer2005,kramer2009,cai2011}. In particular,
\citet{kramer2009} obtain significantly poorer graph recovery results as the
density of the graphs grow. They simulate the corresponding concentration graph
models using the diagonal dominance method, so we have replicated their
experiments but instead using as true models those generated with our proposed
method. The results can be seen in Figure \ref{fig:kramer}, where we have
plotted the true positive rate (also called power by \citet{kramer2009}) and
discovery rates for $p = 100$ and their most dense scenario, $d = 0.25$, when
using matrices simulated with the diagonal dominance method and our proposal.
The different structure learning methods appearing are the same under validation
by \citet{kramer2009}. 
As can be observed, there is significant improvement when
using our method: all of the learning algorithms are close to zero true positive
rate for every sample size when validating on diagonally dominant matrices,
whereas when using matrices obtained via partial orthogonalization, some methods
are able to achieve a true positive rate of $0.45$ approximately. Furthermore,
all true discovery rates are also higher when using matrices simulated by
partial orthogonalization. Importantly, partial least squares regression
performs reasonably good, whereas when only using diagonal dominance one could
erroneously conclude that the method is not well fitted for dense structure
scenarios. This small real example already serves to highlight the practical
application and usefulness of our proposed method.

All the code has been implemented in R~\citep{rcore}. Algorithm
\ref{alg:partortgs} has been implemented directly in C for improved efficiency.
We provide an R package, \texttt{gmat}, with such implementation, which contains
both our method and the dominant diagonal one, available
online\footnote{\url{https://github.com/irenecrsn/gmat}}. We have also
published\footnote{\url{https://github.com/irenecrsn/spdug}} the R scripts used
for generating the data and figures described throughout this section. Thus, all
the above described experiments can be replicated.

\section{Conclusions and future research}\label{sec:conc}
We have proposed a method for generating covariance and concentration matrices
subject to the graphical constraints imposed by an undirected graph. The method
is an alternative to the most commonly employed approach of imposing a dominant
diagonal. As we have shown, the off-diagonal entries in diagonally dominant
matrices suffer from a penalization effect that is worsened as the dimension
increases. We have empirically shown how our method overcomes the structure
recovery difficulties found when validating learning algorithms with the
diagonal dominance method.

We have planned several lines of future research. Since we have obtained very
promising results when using our method in a real validation scenario, it would
be very interesting to explore how other performance measures, and other
structure learning algorithms, are also affected. From the computational point
of view, exploring alternatives to the modified Gram-Schmidt orthogonalization
or taking into account special structures in the graph topology
could reduce the complexity of our approach. Further theoretical results on the
distribution over $\mathbb{S}^{>0}(G)$, for a graph structure $G$, induced by our
method would help to gain insight in properties we have empirically observed,
such as the asymptotic stability in the problem dimension or the relationship
with other matrix distributions such as the hyper Wishart family.

\acks{
This work has been partially supported by the Spanish Ministry of Economy,
Industry and Competitiveness through the Cajal Blue Brain (C080020-09; the
Spanish partner of the Blue Brain initiative from EPFL) and TIN2016-79684-P
projects; by the Regional Government of Madrid through the
S2013/ICE-2845-CASI-CAM-CM project; by Fundación BBVA grants to Scientific
Research Teams in Big Data 2016. Irene Córdoba has been supported by
the predoctoral grant FPU15/03797 from the Spanish Ministry of Education,
Culture and Sports.  Gherardo Varando has been partially supported by a research
grant (13358) from VILLUM FONDEN.}

\bibliography{cor18a}

\end{document}